\documentclass[useAMS,usenatbib]{mn2e}

\usepackage{graphicx}

 \providecommand{\adsurl}[1]{\href{#1}{ADS}}
 
\providecommand{\url}[1]{\href{#1}{#1}}

\newcommand{\merr}{m_{\rm err}}

\newcommand\eq[1]{Eq.~(\ref{#1})}
\newcommand\eqs[2]{Eqs.~(\ref{#1}) and (\ref{#2})}

\newcommand{\be}{\begin{equation}}
\newcommand{\beq}{\begin{equation}}
\newcommand{\ee}{\end{equation}}
\newcommand{\eeq}{\end{equation}}
\newcommand{\bea}{\begin{eqnarray}}
\newcommand{\eea}{\end{eqnarray}}

\newcommand{\rd}{{\rm d}}
\newcommand{\bC}{{\bf \sf C}}
\newcommand{\sC}{{ \sf C}}
\newcommand{\sSigma}{{ \sf \Sigma}}
\newcommand{\bF}{{\bf \sf F}}
\newcommand{\sF}{{ \sf F}}
\newcommand{\sdelta}{{ \sf \delta}}
\newcommand{\bSigma}{{\bf \sf \Sigma}}

\newcommand{\bx}{{\bf x}}
\newcommand{\bv}{{\bf v}}

\newcommand{\br}{{\bf r}}
\newcommand{\brp}{{\bf r'}}
\newcommand{\bri}{{\bf r_{\bf i}}}
\newcommand{\brj}{{\bf r_{\bf j}}}

\newcommand{\hbr}{\hat{\bf r}}
\renewcommand{\(}{\left(}
\renewcommand{\)}{\right)}
\newcommand{\<}{\left<}
\renewcommand{\>}{\right>}
\renewcommand{\d}{\rm d}

\newcommand{\lsim}{\,\raise 0.4ex\hbox{$<$}\kern -0.8em\lower 0.62ex\hbox{$\sim$}\,}

\newcommand{\T}{{^{\rm T}}}
\newcommand{\1}{{^{-1}}}
\begin{document}

\title[Motion of the solar system using SNe
]{
Determining the motion of the solar system relative to the cosmic microwave background using type Ia supernovae
}

\author[Gordon, Land, and Slosar]{Christopher Gordon$^{1}$, Kate Land$^{1}$, and An\v{z}e Slosar$^{1,2}$\\
$^{1}$ Astrophysics Department, University of Oxford, Oxford, OX1 3RH, UK\\
$^{2}$
Berkeley Center for Cosmological Physics, Physics Department and Lawrence Berkeley National Laboratory, \\University of California, Berkeley CA 94720, USA }

\date{}
\maketitle
\begin{abstract}
  We estimate the solar system motion relative to the cosmic microwave
  background using type Ia supernovae (SNe) measurements. We take into
  account the correlations in the error bars of the SNe
  measurements arising from correlated peculiar velocities.  Without
  accounting for correlations in the peculiar velocities, the SNe data
  we use appear to detect the peculiar velocity of the solar system at
  about the 3.5 $\sigma$ level. However, when the correlations are
  correctly accounted for, the SNe data only detects the solar system
  peculiar velocity at about the 2.5 $\sigma$ level.  We forecast that
  the solar system peculiar velocity will be detected at the 9
  $\sigma$ level by GAIA and the 11 $\sigma$ level by the LSST. For
  these surveys we find the correlations are much less important as
  most of the signal comes from higher redshifts where the number density of SNe is insufficient for the correlations to be important.
  \end{abstract}

\section{Introduction}
The cosmic microwave background (CMB) has a 3.4 mK dipole anisotropy \citep{hinshaw06}
which can naturally be explained as being due to the motion of the
solar system with respect to the CMB rest frame \citep{lynlahbur89,strauss92,erdogdu05,loenar07}.  An interesting
consistency check of this is to evaluate the solar system motion
from peculiar velocity surveys (see for example \cite{dalgio00}).  

SNe luminosity measurements provide an accurate probe of peculiar velocities.
 Using observed 
correlations between SNe light curves, we can  estimate the SNe absolute 
magnitudes and thus obtain accurate distance estimates to 
the SNe.  Combined with spectroscopic measurements of the host
galaxies' redshifts, this can be used to estimate the peculiar
velocity of each SNe's host galaxy.  The motion of the solar system  will then
show up as a dipole anisotropy in the SNe derived peculiar
velocities. It is interesting to compare the estimates of
the solar system motion
  from the SNe with those derived from the CMB. If they turn
out to be inconsistent then it may be an indication that there is a
significantly large intrinsic temperature dipole on the CMB surface of
last scattering \citep{turner91,lanpir96}, which could be caused by a double
inflation model \citep{langlois96} for example.

A number of studies have made this comparison
\citep{rieprekir95,Bonvin,jhariekir06}, and a simplifying assumption used
in these studies was that the peculiar velocities of the individual
SNe were uncorrelated with each other.  However, as the peculiar
velocities are caused by variations in the density field, neighbouring SNe
will have correlated peculiar velocities
\citep{wanspertur97,sugsugsas99,Hui:2005nm,bondurgas05,gorlanslo07}. These
correlations will increase the error bars on our peculiar velocity estimate
 as each new SNe measurement
does not represent a completely independent realization of the
velocity field.

In this article we include the correlations of the peculiar velocities
when estimating the motion of the solar system with respect to the
cosmic rest frame.  In Sec.~\ref{example} we give a simple example of
the underestimation of the uncertainty that occurs when correlations
between observations are not taken into account. In Sec.~\ref{method} 
we outline the formalism we use for the SNe correlations, and in
Sec.~\ref{results} we apply the method to SALT calibrated SNe data. In
Sec.~\ref{forecasts} we look at the implications for future surveys.
A summary and discussion of the results is given in
Sec.~\ref{discussion}.

\section{Simple Example of Correlated Errors}
\label{example}
In order to illustrate the effect of correlated errors we consider a
simple example (also discussed in Eq.~5 and its below paragraph of
\citet{Cooray:2006ft}) where we analyse $N$ data points ($x_{i}$) drawn from a
multivariate Gaussian likelihood \beq {\cal L}\propto
|\bC|^{-1/2}\exp(-(\bx-\bmu)^{\rm T}\bC^{-1}(\bx-\bmu)/2) .\eeq
The vector $\bx$ is made up of the data points
($x_{i}$) and each element of the vector $\bmu$ is equal to a
constant, $\mu$. The covariance matrix ($\bC$) has diagonal terms
which are $\sigma^{2}$ and the off-diagonal terms which are
$\rho\sigma^{2}$. That is each data point has correlation $\rho$ with
the other data points.
Suppose that one attempts to estimate the value of $\mu$ and $\sigma^2$ 
from the data and ignores correlations (i.e. assume $\rho$ to be zero). The maximum likelihood
estimators of the mean and variance are then respectively  given by
\beq
m={1\over N} \sum x_{i},
\label{muest}
\quad\quad
s^{2}={1\over N} \sum_{i=1}^{N} (x_{i}-m)^{2}.
\label{nocorrest}
\eeq 
As the likelihood is a multivariate Gaussian distribution \beq
\left<x_{i}\right>=\mu \quad \mbox{and}
\quad\left<(x_{i}-\mu)(x_{j}-\mu)\right>=\sC_{i,j}
\label{nprop}
\eeq where angular brackets denote the expectation value. 
Evaluating \eq{nocorrest} using \eq{nprop}
gives
\beq
 \<m\>=\mu, \quad\quad \<s^{2}\>\approx(1-\rho)\sigma^{2}
\label{nocorrest1}
 \eeq 
in the large N limit, where the
approximation approaches equality for large $N$. 
As can be seen from the above equation, $m$ is an unbiased estimator of the true mean, even when there are correlations in the data. However, as can also be seen, $s^{2}$ is biased by a factor of $(1-\rho)$.

If there where no correlations, the error on $\mu$ can be estimated by
\begin{equation}
  \delta \mu = \frac{s}{\sqrt{N}}\,.
  \label{eq:mubob}
\end{equation}
If there are correlations present then it follows from \eq{nocorrest1} that this estimator will give
\beq
\delta \mu=\sigma\sqrt{\frac{1-\rho}{N}}.
\label{eq:mubob1}
\eeq
Now we find the correct value of $\delta \mu$ to see the effects of ignoring correlations.
For large $N$, the expectation of the  one sigma error on $\mu$ can be approximated
\beq
\delta\mu \approx\sqrt{(\sF^{-1})_{\mu,\mu}}
\eeq
where $(\sF^{-1})_{\mu,\mu}$ is the $(\mu,\mu)$ component of the inverse of the Fisher matrix.
For a multivariate Gaussian likelihood the Fisher matrix is given by  (see for example \citet{tegtayhea96})
\bea
\bF_{\alpha,\beta} &\equiv&
-\left< \frac{\partial^2\ln{\mathcal L}}{\partial p_\alpha \partial p_\beta}\right>
\\ &=&\bmu_{,\alpha} \bC^{-1} \bmu\T_{,\beta} + 
\frac{1}{2} {\rm Tr}\left( \bC\1 \bC_{,\alpha} \bC\1 \bC_{,\beta}\right)
\label{fisher}
\eea where $\alpha$ and $\beta$ run over the different model
parameters which are being estimated ($\mu$ and $\sigma$, with $\rho$
assumed known, in the current example). \eq{fisher} gives the true error on the 
value of $\mu$, and in our example this is
\beq
\delta\mu_{{\rm corr}} = \sigma\sqrt{\frac{1+(N-1)\rho}{N}}.
\label{dmu}
\eeq
Comparing \eqs{eq:mubob1}{dmu} we see that if the data are 
correlated ($\rho>0$) but correlations are
 neglected then one would underestimate the uncertainty on $\mu$. One would 
 overestimate the error if the data is
anti-correlated, although we note that this case is restricted as for the covariance matrix 
to be positive definite, $\rho>-(N-1)^{-1}$.

As was shown in a earlier studies \citep{Hui:2005nm,nehuco07,gorlanslo07}, an
analogous underestimation of the error happens if the correlations in
low redshift SNe are not accounted for when using them in a sample to
estimate the dark energy equation of state, $w$. In this article we
show that there is also an underestimation in error on the motion of
our solar system when the correlations in the SNe are not accounted
for.

\section{Method}
\label{method}
The luminosity distance, $d_L$, to a SN at redshift $z$, is defined such that
\[
{\mathcal F} = \frac{\mathcal L}{4\pi d_L^2}
\]
where $\mathcal F$ is the observed flux and $\mathcal L$ is the SN's intrinsic
luminosity. Astronomers use magnitudes, which are related to 
the luminosity distance (in megaparsec) by
\be
 m - M = 5\log_{10} d^{obs}_L + 25\label{dlobs},
\ee
where $m$ and $M$ are the apparent and absolute magnitudes
respectively.  In the context of SNe, $M$ is a ``nuisance parameter''
which is completely degenerate with $\log(H_0)$ and is marginalised
over.  For a Friedmann-Robertson-Walker Universe the predicted
luminosity distance is given by \be d_{L}(z)= (1+z)\int_{0}^z
\frac{{\rd}z'}{H(z')}\label{dlth} \ee (taking $c=1$), where
$H$ is the Hubble parameter.  In the limit of low redshift this
reduces to $d_{L}\approx z/H_{0}$.

Given very stringent limits on the curvature of the universe, we can
safely work within the assumption of a flatness as the allowed
curvature would not play any role at the scales of interest. In this
case,  the effect of a peculiar velocity (PV) leads to 
a perturbation in the luminosity distance ($\delta d_{L}$)  given by
~\citep{sasaki87,sugsugsas99,pynbirk95,bondurgas05,Hui:2005nm}
\begin{equation}
\frac{\delta d_{L}}{d_{L}} = \hbr \cdot \(\bv-\frac{(1+z)^{2}}{ 
H(z) \: d_{L}}[\bv-\bv_O]\)
\label{ddod}
\end{equation}
where $\br$ is the position of the SN, 
and $\bv_O$ and $\bv$ are the peculiar velocites of the observer and SN repectively. 
In the limit of low redshift, 
$\delta d_{L}\approx  \hbr\cdot [\bv_{O}-\bv]/H_{0}$.
This demonstrates how a SNe survey that measures $m$ and $z$ can 
estimate the projected PV field. We now relate this to the cosmology.

The projected velocity correlation function, 
$\xi(\br,\brp)\equiv \< (\bv(\br)\cdot\hbr)( \bv(\brp)\cdot\hbr')\>$, 
must be rotationally invariant, and therefore it can be 
decomposed into a parallel and perpendicular 
components~\citep{Gorski,Groth,Scott} :
\be
\xi(\br,\brp)=\sin\theta \sin\theta' \xi_{\perp}(\Delta r,z,z')+\cos\theta
\cos\theta'\xi_{\parallel}(\Delta r,z,z')
\ee
where $\Delta \br \equiv \br-\br'$, $\Delta r=|\Delta \br|$,
$\cos\theta\equiv\hbr\cdot\Delta \hbr$, 
and $\cos\theta'\equiv\hbr'\cdot\Delta \hbr$. In linear theory, 
these are given by~\citep{Gorski,Groth,Scott}:
\be
\xi_{\parallel,\perp}
= D'(z)\:D'(z')\: \int_{0}^{\infty} \frac{\d k}{2\pi^{2}}P(k) 
K_{\parallel,\perp}(k r)\label{xith}
\ee
where  for an arbitrary variable $x$, 
$K_{\parallel}(x)\equiv j_0(x)-\frac{2j_1(x)}{x}$, $K_{\perp}(x)\equiv j_1(x)/x$. 
$D(z)$ is the growth function, and derivatives are with respect to conformal 
time. $P(k)$ is the matter power spectrum which can be evaluated either numerically
(e.g.\ CAMB \cite{lewcha99}) or using analytical approximations \citep{eishu97}.

The above estimate of $\xi(\br,\br')$ is based on linear theory.
On scales smaller than about 10$h^{-1}$Mpc, nonlinear contributions dominate.
These are usually 
modeled as an uncorrelated term which is independent of redshift, 
often set to $\sigma_{v}\sim 300$ km/s. 
Comparison with N-body simulations \citep{silberman01} indicate that this is an
 effective way of accounting for the non-linearities.
Other random errors that are usually considered 
are those from the lightcurve fitting ($\merr$), 
and  intrinsic magnitude scatter ($\sigma_m$) .
It is just these three errors that are usually included in the 
analysis of SNe.

 The residual deviations of luminosity distance from the homogeneous
expansion can be packed into a data vector
\begin{equation}
\left(\frac{\delta d_{L}}{d_L}\right)_i=\frac{d_L^{obs}(i)-d_L(z(i))}{d_L(z(i))},
\end{equation}
 whose covariance matrix (from the correlated PVs) is given by
 \be \sC_v(i,j)=\left( 1 - \frac{(1+z)^2}{H\: d_L}\right)_i
\left( 1 - \frac{(1+z)^2}{H\: d_L}\right)_j \xi(\bri,\brj)\label{Cv},
\ee
while the standard uncorrelated random errors are given by
\be
\sigma(i)^{2}=\left({\ln(10)\over 5}\right)^{2}
(\sigma_{m}^{2}+\merr(i)^{2})+\left( 1 - \frac{(1+z)^2}{H\:
    d_L}\right)_i^2 \sigma_v^2. \label{diag}
\ee
Some example plots of $\sC_{v}$ were given in \citet{gorlanslo07}.
The likelihood is then 
\be
{\cal L} = (2 \pi)^{-N/2} |\bSigma |^{-1/2}\exp\left({-{1\over 2}{
      \mathbf{\Delta}}^{T}\bSigma^{-1}{\mathbf{\Delta}}} \right)
\label{lik}
\ee
 where 
 \be
 \sSigma(i,j) =\sC_{v}(i,j)+\sigma(i)^{2}\sdelta_{ij}
 \label{Sigma}
 \ee
  and 
\be
\Delta_i=\left(\frac{\delta d_{L}}{d_L}\right)_i-\(\frac{(1+z)^{2}}{ 
H(z) \: d_{L}}\)_{i}\hbr_i \cdot \bv_O.
\label{x}
\ee
We now proceed to find constraints on the observer velocity $\bv_O$. We
assume a standard $\Lambda$CDM cosmology and impose Big Bang
Nucleosynthesis (BBN) prior $\Omega_{b} h^{2} \sim{\mathcal
  N}(0.0214,0.002)$~\citep{bbn}, and a Hubble Space Telescope (HST)
prior $h\sim {\cal N}(0.72,0.08)$~\citep{hst}. These two priors remove
models that are wildly at odds with standard cosmological probes, but
do not unduly bias results towards standard cosmology. The likelihood
has almost negligible dependence on $n_s$, and to keep it in a range
consistent with CMB and large scale structure estimates we give it a
uniform prior $n\in[-0.9,1.1]$.

We parameterize the solar system peculiar velocity as a magnitude
($v_O$) and direction in galactic coordinates $(l,b)$.  The prior on
$(l,b)$ was assumed to be uniform on the sphere, i.e flat on $l$ and
$\cos(b)$. The prior on $v_{O}$ was set to be uniform. 

\begin{figure}
\centerline{\includegraphics[width=10cm]{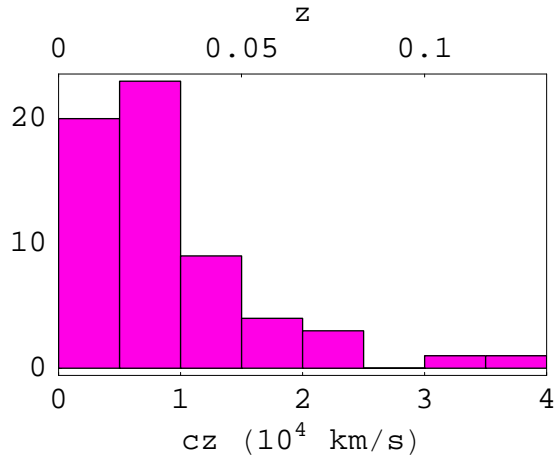}}
\caption{Heliocentric distribution of redshifts for the low $z$ sample.
\label{histo}}
\end{figure}

\begin{figure}
\centerline{\includegraphics[width=10cm]{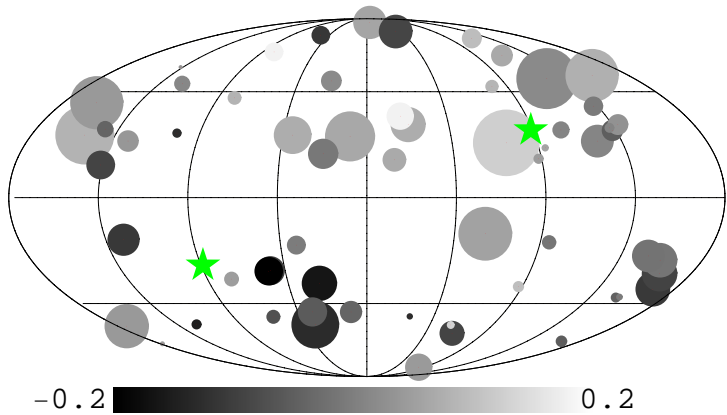}}
\caption{Positions  of the SNe for the low $z$ sample in galactic coordinates.
The size of the disk is inversely proportional to the redshift. The color of the disk is related to the relative luminosity distance error ($\delta d_{L}/d_{L}$). 
The stars indicate the direction of the CMB dipole.
\label{pos}}
\end{figure}

We use a SALT \citep{guy05} calibrated low redshift SNe data
set\footnote{Obtained from
  http://qold.astro.utoronto.ca/conley/bubble/} with heliocentric
redshifts in the range $cz \in [2278,37163]$km/s, ($z\in
[0.0076,0.124]$).  A histogram of the 61 redshifts is shown in
Fig.~\ref{histo} and the sky positions of the data are shown in
Fig.~\ref{pos}.
We also used the higher redshift 71 SNe from the SNLS data
set.\footnote{Obtained from http://snls.in2p3.fr/conf/papers/cosmo1/}

The SALT calibration involves the additional parameters
($\alpha$, $\beta$), which account for the shape/luminosity and
colour/luminosity relations of SNe. These and the other parameters
$(\Omega_m,\sigma_8,\sigma_v,\sigma_m,M)$
 are all
given broad uniform priors. We use the standard Markov Chain Monte
Carlo (MCMC) method to generate samples from the posterior
distribution of the parameters~\citep{cosmomc}.  Convergence was
checked using multiple chains with different starting positions, and
also the $R-1$ statistic ~\citep{mcmc}.  We also checked that the
estimated posterior distributions reduced to the prior distributions
when no data was used.
The analysis was checked with two completely independent codes and
MCMC chains.

Additionally, we looked at the combination of the SNe observations
with the WMAP data of the CMB \citep{WMAP06} (with the usual CMB priors in this case). We stress that we do not use the WMAP dipole
information, but rather just the $\ell>1$ information, which when
combined with the SNe has the effect of constraining matter density
and the amplitude of matter fluctuations.

\section{Results}
\label{results}
\begin{figure}
\centerline{\includegraphics[width=8cm]{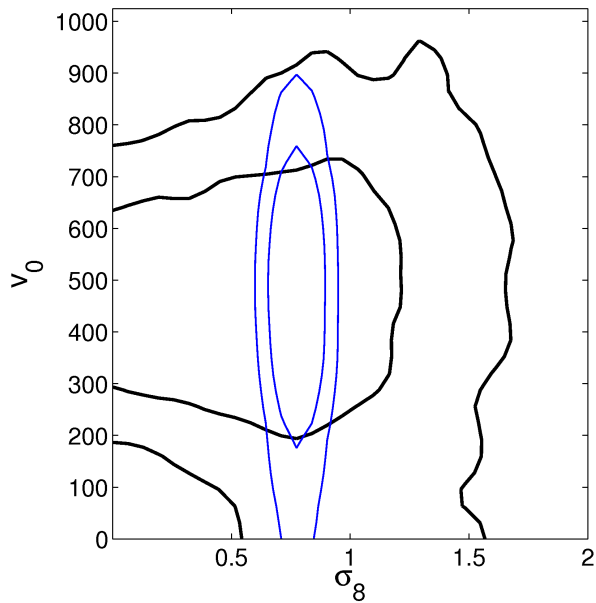}}
\caption{Marginalized one and two sigma contours for the magnitude of
the solar system peculiar velocity $(v_O)$ vs the dispersion of the matter density field smoothed on scales of 8$h^{-1}$Mpc ($\sigma_{8}$). The thick (black) contours are for SNe data with BBN and HST priors. The thin (blue) contours are for SNe with a WMAP prior.
}
\label{v02D}
\end{figure}

In Fig.~\ref{v02D} we plot the marginalized probability contours for
$v_O$ and $\sigma_{8}$, where we see that higher values of
$\sigma_{8}$ have broader contours on $v_O$. This is because a
larger $\sigma_{8}$ implies more correlations between the SNe peculiar
velocities and so less of a reduction in the errors due to averaging
effects. The contours for when WMAP is included are also plotted.

\begin{figure}
\centerline{\includegraphics[width=10cm]{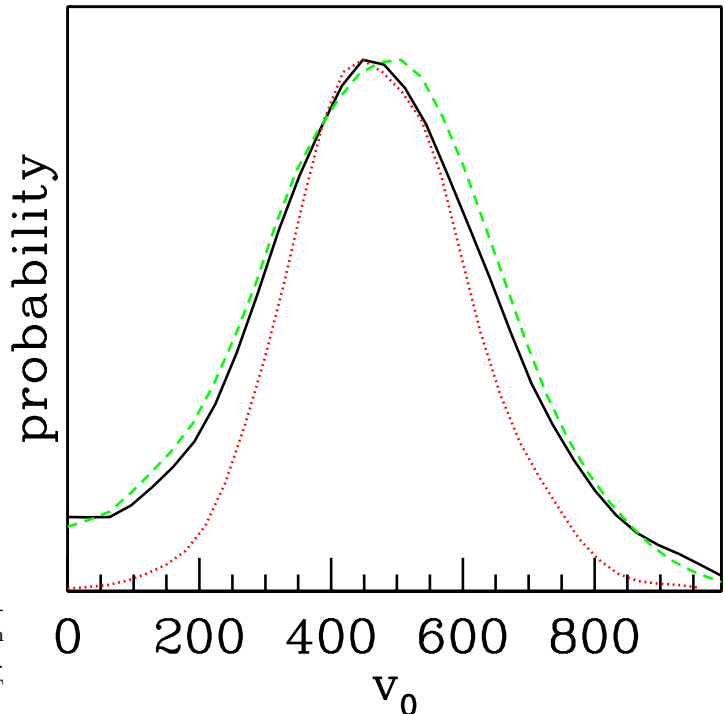}}
\caption{ 
Marginalized probability distributions for the magnitude of the solar system peculiar velocity, $v_O$. The dotted line is for when the correlations in the SNe peculiar velocities are not accounted for. The solid line is with correlations and the dashed line is with correlations and WMAP temperature data included to reduce the uncertainty in  the cosmological
parameters.
}
\label{v0}
\end{figure}

\begin{figure}
\centerline{\includegraphics[width=10cm]{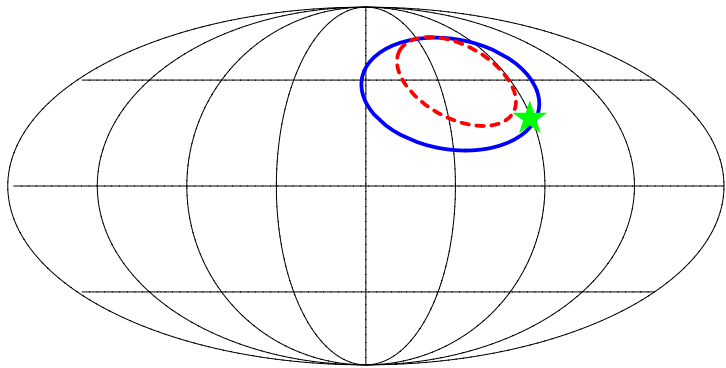}}
\caption{ 
\label{dipole}
One sigma contours for the direction of the solar system velocity. The cases plotted are when correlated (solid) and uncorrelated (dashed) SNe peculiar velocities are used. The star shows the direction as determined by the CMB.
}
\end{figure}

In Fig.~\ref{v0} a marginalized probability distribution is plotted
for $v_O$, and we see that the uncertainty on $v_O$ increases
dramatically when the correlations are accounted for.  It can also be
seen that adding WMAP data has a negligible effect. As seen from
Fig.~\ref{v02D}, this arises because the WMAP data constrains
$\sigma_{8}\approx0.8$, which happens to lie on an approximately average value
for the $v_O$ uncertainty.  In Fig.~\ref{dipole} the one sigma
confidence intervals are plotted for the direction of the solar system
motion. As can be seen, not accounting for the correlations
underestimates the uncertainty by about a factor of 2.

Also, Figs.~\ref{v0} and \ref{dipole} shows that the SNe data are
consistent with the CMB dipole estimate of $ (v_O,l,b)=(369\pm 3 {\rm
  km/s}, 263.86\pm 0.04^\circ, 48.24\pm 0.10^\circ)$
\citep{hinshaw06}.  In Table~\ref{resultstab} the mean and
uncertainties of the solar system peculiar velocity are given. 
We find
that when the correlations are not included, the estimate of $v_O$ is
about 3.5 standard deviations from zero. While if the correlations are
accounted for then $v_O$ is only about 2.5 standard deviations from
zero. One can convert the number of standard deviations of the
detection into upper bounds on the Bayesian odds ratio
\citep{gortro07}. Without correlations the odds, from SNe data, of
$v_O$ being non-zero appear to be at best 119:1. While if the correlations are
accounted for then the odds are at best only 7:1.
In Table~\ref{resultstab} we also give an estimate for the local group motion which was obtained by subtracting the solar system velocity relative to the local group \citep{yahtamsan77}.
\begin{table}\centering
\begin{tabular}{l||ccc}
                                      &$ l$                   &$ b $              & $ v_O$ (km/s)        \\     \hline 
Solar System uncorrelated          &  $238\pm 26^\circ$  & $ 45\pm14^\circ $ & $475 \pm 134$     \\
Solar System correlated            &  $234\pm 44^\circ$  & $39\pm21^\circ$  & $468\pm 186$   \\
Local Group uncorrelated         &  $260 \pm 14^\circ$  & $ 32 \pm 11^\circ $ & $697 \pm 137$\\
Local Group correlated          &  $257 \pm 24^\circ$  & $29 \pm 16^\circ$  & $690 \pm 201$   
\end{tabular}
\caption{
\label{resultstab}
The mean and standard deviation for the estimate of the solar system and local group velocity from current SNe data. The results for both the correlated and uncorrelated peculiar velocities  are shown.
}
\end{table}

\section{Forecasts}
\label{forecasts}

In this section we forecast constraints on the motion of the solar
system from GAIA (the ``super-Hipparcos'' satellite) and LSST.  For
GAIA, based on the simulations by \citet{beleva02}, we generate a
sample of 6,317 SNe distributed over the full sky with $z <0.14$. For
LSST we generate 30,000 SNe distributed over the full sky with $z
<0.3$ \citep{wanpinzha05}.  We weighted the distribution of SNe by
$\cos(b)z^2$ to account for the volume in spherical coordinates, that is we keep
the density constant with $z$. As
our fiducial model we took $\{v_O, l, b, \Omega_m, \Omega_b, h, n_s,
w, \sigma_8,\sigma_v,\sigma_m,m_{err}\}=
\{369,264^\circ,48^\circ,0.3,0.041,0.72,0.96,-1,0.85,300,0.1,0.1\}$.
We do not include the SALT calibration parameters $(\alpha,\beta)$ in
the forecast, but we have checked using forecasts for the data sets
used in Sec.~3 that this does not have a significant
effect. 
In order to use the Fisher matrix (see \eq{fisher}), we consider the function
\be
d = d_L^{\rm obs}\:10^{M_0/5}
\ee
where $d_L^{\rm obs}$ is given by Eq.(\ref{dlobs}). 
The expectation value vector has each element given by
\be
 \left<d\right> =
10^{M_0/5}\left(d_L^{\rm th}+\hat{\br}\cdot\bv_O\:\frac{(1+z)^2}{H(z)}\right),
\ee
with $d_L^{\rm th}$ given by Eq.(\ref{dlth}). The covariance matrix $(\bC)$ is 
as before, Eq.(\ref{Sigma}), but with the extra 
factor $\left(10^{M_0/5}\right)^2 d_L(i) d_L(j)$. Note the reason for the slightly different function of the data compared to \eq{x} is so that the
data vector does not depend on any of the parameters.

In Table~\ref{forecasttab} we present our main forecast 
results. 
\begin{table}\centering
\begin{tabular}{l||ccc}
                                      &$ l$                   &$ b $              & $ v_O$ (km/s)        \\     \hline 
GAIA Uncorrelated PVs          &        $8^\circ$             &      $5^\circ$                &      36                 \\
LSST Uncorrelated PVs          &        $7^\circ$             &      $5^\circ$                &      32                 \\
GAIA Correlated PVS             &      $10^\circ$              &   $6^\circ$               &       42                 \\
LSST Correlated PVS             &      $8^\circ$              &   $5^\circ$               &       34                 \\
\end{tabular}
\caption{
\label{forecasttab}
The forecasted marginalised standard deviation for the estimate of the solar system peculiar velocity from the future SNe surveys GAIA and LSST, where SNe peculiar velocities (PVs) are treated as uncorrelated and correlated. 
}
\end{table}
As can be seen there is a dramatic improvement in the constraints
compared to current data. Also, unlike current data, taking into
account the correlations does not have a large effect. This is because 
most of the constraining power for GAIA and LSST comes from
higher redshifts, where the peculiar velocity errors are negligible
compared to the other types of error.  This can be understood as
follows.  From Eq.~\ref{Cv} we see that the 
error induced by peculiar coherent velocity flows drops
as $1/z$, and thus for high enough redshift ($z>0.015$ for a typical experiment) 
they become unimportant compared to the redshift independent 
measurement errors in \eq{diag}. In this
limit the weakening of the dipole signal, that drops as $1/z$ in Eq.(\ref{x}), is
exactly compensated by the number of supernova in a redshift slice,
which increases as $z^2$, for a volume-weighted survey. The signal to
noise is therefore low and increasing at low redshift, tailing
off to a constant value at redshifts at which peculiar velocities
become unimportant.

\begin{figure}
\centerline{\includegraphics[width=8cm]{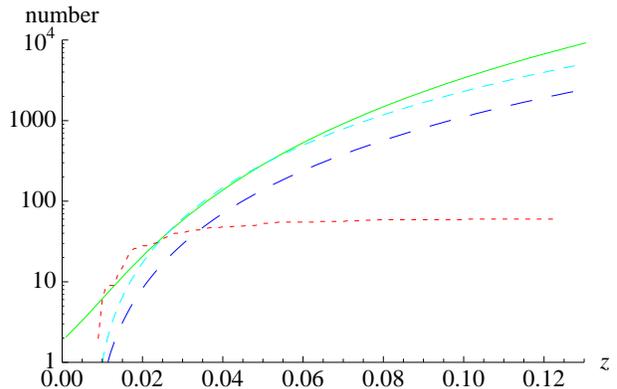}}
\caption{A plot of the lower bound on the number of SNe needed at
or with redshift less than $z$ for correlations in the peculiar velocities to be important
(solid line). Also plotted are the number of SNe, with a redshift less than or equal to $z$, for the currently available low redshift SNe
(dotted line) and for volume weighted GAIA (short dashed) and 
LSST (long dashed) surveys.
\label{whencorrneeded}
}
\end{figure}

As a rough measure of whether the correlated
error will be important, at a redshift $z$, we look at the ratio
N=$\sigma^{2}/\bC_{v}$ where $\bC_{v}$ is evaluated using \eq{Cv} with
two SNe both at redshift $z$ and 90$^{\circ}$ apart. This is
effectively the ratio of the measurement error on the SN
luminosity to its covariance with a typical SN in the
dataset. \eq{dmu} tells us that this is approximately equal to the number of SNe for which 
the covariance will become important to our error estimates, and we plot this in
Fig.~\ref{whencorrneeded}.
As can be seen this simple estimate is in agreement with the more
complete analysis: current data are effected by correlated errors much
more so than GAIA and LSST, which will be practically uneffected by
correlations. In other words, under the assumption of a volume weighted 
redshift distribution for the SNe, GAIA and LSST can rely on the much
higher redshift data to constrain our peculiar motion, as these are
 considerably less affected by PV correlations. 

\section{Discussion}
\label{discussion}

To summarize, we have used SALT calibrated SNe data to estimate the motion of he solar system.
As seen from Table~\ref{resultstab}
the error bars are
under-estimated by about 50\% if the correlations in the peculiar
velocity are not accounted for.

We now compare our findings to previous published results.
In \citet{Bonvin} they used 44 SALT calibrated SNe. They only allowed $ \{l,b,v_O\}$ to vary and all the other parameters where fixed to standard values. They did not account for correlations in the peculiar velocities. They found $v_{0}=405\pm192$km/s which is compatible with our result. 

In \citet{jhariekir06} they used 69 SNe with $z\in\{0.005,0.025\}$. They also only allowed $ \{l,b,v_O\}$ to vary. Additionally, they used MLCS2k2 to calibrate the data, rather than the SALT method.
They also did not account for the correlations in the peculiar velocities.
 They evaluated the motion of  the local group and found
$\{l,b,v_O\}=\{258\pm18^{\circ},51\pm12^{\circ},541\pm75{\rm km/s} \}$. 
Our  local group velocity results, which are listed in  Table~\ref{resultstab}, are compatible with those of \citet{jhariekir06}  but, even when we don't take into account the correlations in the peculiar velocities, our error on the magnitude of $v_{0}$ is about 80\%
 larger than that of  \citet{jhariekir06} study. This is due to several factors. They had a lower redshift limit than us: 0.005 vs. 0.0076. One of the reasons we did not go to such a low redshift is that the peculiar velocities (including the motion of our solar system) become of order the Hubble expansion. This means that the motion of our solar system has a  high signal (hence the low error bars obtained by \citet{jhariekir06}) but one can no longer use \eq{ddod} to evaluate the effects of peculiar velocity on the luminosity distance. It would be possible to use a higher order version of \eq{ddod} but this was not done by \citet{jhariekir06} and so their
results will have an additional unreported systematic error due to the induced luminosity change being calculated incorrectly. Also, the extra very low redshift SNe that were used are dominated by SNe that are too close together to model the correlations in the peculiar velocity using  linear theory, \eq{xith}. Overall, our results are the only ones that take into account the correlations in peculiar velocities and account for the uncertainties in the cosmological and calibration parameters.

We also made forecasts for the GAIA and LSST surveys, assuming 
volume weighting for the redshift distribution. We found that
the error bars will be about 4 times smaller than those of current
data, but still not competitive with those from the CMB by a factor of 
$\sim 10$ (assuming that the CMB dipole 
is due to our local motion). Also, for GAIA and 
especially LSST we found that correlations
had little effect as most of the signal came from higher redshifts
where the correlations are almost negligible for the sample sizes
considered.

In future work, we plan to test the techniques we have used in this paper
against simulated SNe surveys generated from N-body
simulations. These will test the assumptions that
go into our data modeling - most importantly the effect of
non-linearities \citep{haugboelle06}.

In \cite{watfeld07} it was shown that the large scale properties (bulk
flow and shear \citep{kaiser91,jafkai95}) of the peculiar velocity field derived from a low
redshift sample of 73 SNe was consistent with the bulk flow and shear
of the velocity field derived from the SFI, ENEAR, and SBF surveys. 
It would be interesting to combine all these surveys together (with SFI replaced by SFI++ \citep{springbob07}) to estimate
the solar system velocity with respect to the CMB and put robust limits on the intrinsic CMB dipole.

\section*{Acknowledgments}
We thank Francesco Calura, Mark Sullivan, Licia Verde, and Joe Zuntz
for helpful discussions.  CG is funded by the Beecroft Institute for
Particle Astrophysics and Cosmology, KL by a Glasstone research
fellowship, and AS by Oxford Astrophysics and the
Berkeley Center for Cosmological Physics.

\bibliographystyle{mn2e_eprint}

\bibliography{bib}

\end{document}